# Scalable Quantum Photonics with Single Color Centers in Silicon Carbide

*Marina Radulaski\*[†], Matthias Widmann\*[§], Matthias Niethammer[§], Jingyuan Linda Zhang[†], Sang-Yun Lee[§∂], Torsten Rendler[§], Konstantinos G. Lagoudakis[†], Nguyen Tien Son[‡], Erik Janzén[‡], Takeshi Ohshima[÷], Jörg Wrachtrup[§], Jelena Vučković [†]*

[†] E. L. Ginzton Laboratory, Stanford University, Stanford, California 94305, United States

[§] 3rd Institute of Physics, IQST and Research Center SCOPE, University of Stuttgart, 70569 Stuttgart, Germany

[∂] Center for Quantum Information, Korea Institute of Science and Technology (KIST), Seoul, 02792, Republic of Korea

[‡] Department of Physics, Chemistry and Biology, Linköping University, SE-58183 Linköping, Sweden

[÷] National Institutes for Quantum and Radiological Science and Technology (QST), Takasaki, Gunma 370-1292, Japan

ABSTRACT

Silicon carbide is a promising platform for single photon sources, quantum bits (qubits) and nanoscale sensors based on individual color centers. Towards this goal, we develop a scalable array of nanopillars incorporating single silicon vacancy centers in 4H-SiC, readily available for efficient interfacing with free-space objective and lensed-fibers. A commercially obtained substrate is irradiated with 2 MeV electron beams to create vacancies. Subsequent lithographic process forms 800 nm tall nanopillars with 400-1,400 nm diameters. We obtain high collection efficiency, up to 22 kcounts/s optical saturation rates from a single silicon vacancy center, while preserving the single photon emission and the optically induced electron-spin polarization properties. Our study demonstrates silicon carbide as a readily available platform for scalable quantum photonics architecture relying on single photon sources and qubits.

KEYWORDS: color centers, silicon carbide, photonics, spintronics, nanopillars, spin-qubits.

TABLE OF CONTENT GRAPHIC

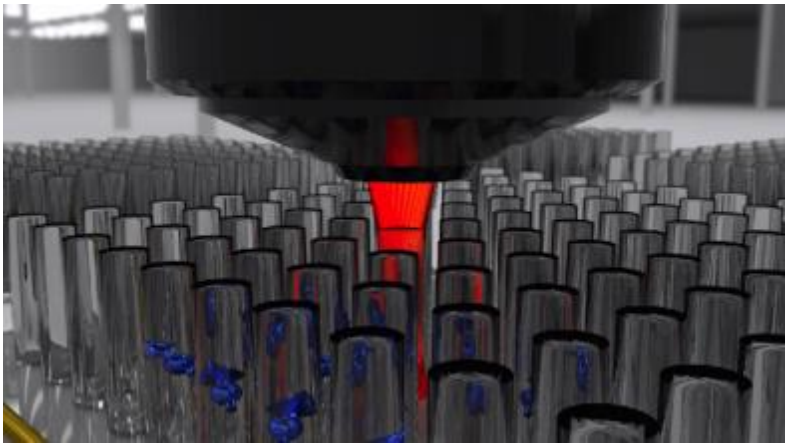

Silicon carbide (SiC) has recently emerged as a host of color centers with exceptional brightness[1] and long spin coherence times,[2-5] much needed for the implementations of solid-state quantum bits and nanoscale magnetic sensors.[6] In addition to a favorable set of physical properties, such as the wide band gap and high index of refraction, this material also benefits from the decades of utilization in power electronics[7] – wafer scale substrates are available for purchase, doping and various etching processes[8-11] are at an engineer's disposal. This makes SiC advantageous over diamond and rare-earth doped crystals for industrial applications in quantum computing and sensing, as well as for integration of point defects with electronics and CMOS devices. Electron spins of various SiC color center ensembles have been coherently controlled to date.[3,12,13] We focus our attention on individual silicon vacancy ($V_{Si}$) centers in 4H-SiC, whose electron spin was recently employed as room temperature qubit,[2] the first one at the single defect level in SiC. This system features spin coherence times of at least 160 μs, and with a potential for improvement through isotopic purification. To bring this system closer to the applications, we address $V_{Si}$/4H-SiC platform from the points of scalability, device footprint, optical interface and collection efficiency, while preserving optically induced electron-spin polarization properties.

Our platform consists of a 2D array of simultaneously fabricated nanopillars in electron beam irradiated 4H-SiC. The nanopillars feature 0.16-2 $\mu m^2$ footprint and efficient signal collection from single color centers into a free space objective of NA = 0.65. Such numerical aperture is also available among lensed fibers,[14] as a potential alternative interface. In comparison, a previous approach has relied on the non-scalable fabrication of solid immersion lens (SIL) interface.[2] Here, focused ion-beam milling with heavy ions is used, and an intense combination of wet and reactive ion etching needs to be performed to remove radiation damage in the SiC lattice. A footprint of an individual SIL is ~1,000 $\mu m^2$. In another approach, a bulk wafer was irradiated to generate a 2D array of color center sites.[15] Both of these approaches relied on collection into high numerical aperture (NA = 1.3) oil immersion lens, in a direct contact with the substrate. Vertically directed waveguiding in nanopillars allowed us to observe up to 22 kcounts/s (kcps) collection from single color centers with half the NA value. This is comparable to the best previous result of 40 kcps achieved in a non-scalable system,[2] and a threefold increase over 7 kcps result in a scalable array in bulk.[15] The use of NA = 0.95 air objective lens would result in a few-fold increase in collection efficiency, as discussed in Supporting Information. We also observe 2-4% relative intensity in optically detected magnetic resonance (ODMR), comparable to the previous result.[2] Finally, the preservation of optical and spin properties of $V_{Si}$ centers makes this platform applicable for wafer-scale arrays of single photon sources and spin-qubits.

**Fabrication.** We develop a scalable nanofabrication process, based on electron irradiation and electron beam lithography, which produces tens of thousands of nanopillars per squared millimeter in 4H-SiC substrate that incorporates $V_{Si}$ color centers. Fig. 1a illustrates the process flow. The initial sample is a commercially available high purity semi-insulating 4H-SiC crystal, irradiated with 2 MeV electrons at a fluence of $10^{14}$ cm$^{-2}$ at room temperature. For the purposes of hard etch mask, 300 nm of aluminum was evaporatively deposited onto the sample. Microposit ma-N negative tone resist was spun over it for patterning of nanopillars in 100 keV JEOL (JBX-6300-FS) electron beam lithographic system. Aluminum hard mask was etched in $Cl_2$ and $BCl_3$ plasma, and the pattern was transferred to silicon carbide in an $SF_6$ plasma process, both performed in the Versaline LL-ICP Metal Etcher. The mask was stripped in KMG Aluminum Etch 80:3:15 NP.

The pattern consists of 10 × 10 arrays of nanopillars with 5 μm pitch and diameters varied between 400 nm and 1.4 μm. The profile features 800 nm tall cylinder whose upper third has a shape of a cone frustum, reducing the top diameter by 120 nm. Fig. 1b-c show the SEM images of the fabricated structures.

**Collection efficiency.** We model emission of color centers placed in pillars using Finite-Difference Time-Domain method. Index of refraction distribution is made according to the profile of fabricated pillars ($n_{SiC}$ = 2.6, $n_{air}$ = 1) and continuous wave dipole excitation at 918 nm is positioned within. We consider nanopillars of 400 – 1,400 nm diameter with excitation at the central axis of the pillar ($r = 0$, $z = 400$ nm) with in-plane ($E_r$) and vertically oriented ($E_z$) dipole components. The orientation of $V_{Si}$ dipole in our samples is along main crystal axis which forms $\theta = 4°$ angle with $z$-axis, and therefore we add the components to obtain total collected field $E = E_z \cos\theta + E_r \sin\theta$ (Fig. 2a). We evaluate the fraction of collected emission by a two-step analysis. First, we monitor the power that penetrates the walls of the simulation space and evaluate the fraction of light that is emitted vertically upward (Fig. 2b). Next, we analyze the far field image (Fig. 2c) and obtain the power fraction of vertically emitted light that is collected within a given numerical aperture (NA). Due to the dominant contribution of the dipole-component along the $z$-axis, which is also noticeable in Fig. 2a, we consider only vertical excitation $E_z$ for additional analyses (details in Supporting Information). In an extensive search for the optimal dipole-emitter position for each pillar diameter, we learn that color center positioning closer to the bottom of the pillar is advantageous for collection efficiency. We also run this analysis for unpatterned silicon carbide, in order to estimate the expected collection efficiency increase achieved through nanopillar waveguiding, which yields 1-2 orders of magnitude improvement. The results are presented in Fig. 2d-e showing favorable collection efficiency of 600 nm and 1,000 nm diameter nanopillars.

**Single photon emission.** Photoluminescence (PL) measurements obtained in a home-built confocal microscope at room temperature are shown in Fig. 3. Light from a single-mode laser diode operating at 730 nm was sent to an air objective (Zeiss Plan-Achromat, 40x/0.65 NA), in order to excite the defects. The PL signal was collected with the same objective and sent through a dichroic mirror and subsequently spatially filtered by a 50 μm diameter pinhole. The photons emitted by the defects were spectrally filtered by a 905 nm long pass filter and finally collected with two single photon counting modules in a Hanburry-Brown-Twiss configuration, or, alternatively, with a spectrometer (details presented in Supporting Information). The density of color centers is too high to resolve single defects in the bulk crystals with no fabricated structures, however, combined with the nanopillars, a spatial separation of single defects is possible, simultaneously providing a moderate yield of defects placed in the pillars, as shown in Fig 3a. Approximately 4 out of 100 pillars host single $V_{Si}$ defects which desirable lateral and vertical placement, and show an enhanced photon count-rate compared to defects without photonic structure[15]. For each investigated $V_{Si}$ defect a laser power dependent saturation curve of the PL was collected as shown in Fig. 3b. The background was obtained by collecting the light from an empty pillar nearby and subtracted from the defects emission. The influence of the diameter was experimentally examined by a statistical analysis of the maximum PL intensity as a function of the diameter. The maximum PL was obtained by fitting the background-corrected saturation curve with a power law: $I(P) = \frac{I_s}{1+\alpha/P}$, where $I$ is the PL intensity as a function of the excitation power $P$, $\alpha$ is excitation power when saturated and $I_s$ denotes the saturation photon count-rate. The result of the fit parameter $I_s$ is plotted as a function of the diameter in Fig. 3c, and show a qualitative match to the modeled collection efficiency trend. The error-bars were obtained by the error from the least-square fits. The maximum PL from a single defect was found in the pillars with 600 nm diameter, showing $I_{max} = 22$ kcps, which is an increase of factor of 2-3 compared to 5-10 kcps for a single $V_{Si}$ defect without photonic structures reported in literature[2,15,16]. This increase is smaller than expected from the model, which is attributed to the imperfect fabrication profile. A typical room temperature PL spectrum is shown in Fig. 3d, while Fig. 3e shows a typical second order autocorrelation measurement with a clear single photon emission characteristic.

**Single spin optically detected magnetic resonance at room temperature.** In order to address the single spins, a radio-frequency (RF) signal was generated by a Rohde and Schwarz vector signal generator, subsequently amplified and applied to the $V_{Si}$ in pillars through a 25 μm copper wire, which was spanned over the sample. Detailed information about the experimental methods for ODMR are described in a previous publication.[2] Continuous wave (cw) ODMR is observable under cw laser excitation where spin-dependent transitions over a metastable state lead to a spin polarization into the magnetic sublevels. At the same time RF is swept over the spin resonance. When the RF signal is resonant to the Larmor-frequency, the spin state is changed and the transition rates over the metastable state is altered, hence a change in PL intensity can be detected. About 10% of the single $V_{Si}$ defects which show enhanced emission also showed clear spin imprints, as illustrated in a typical ODMR measurement at a zero magnetic field plotted in Fig. 3f, which is in agreement with previous studies.[2,15] $V_{Si}$ defects in 4H-SiC can originate from two inequivalent lattice sites.[17] Among them, the so-called V2 center can be identified by their 70 MHz zero-field splitting[18] of the ground state spin of $S = 3/2$. The center position at 70 MHz of a broadened ODMR spectrum as shown in Fig. 3f confirm that the found $V_{Si}$ defects showing ODMR signals are V2 centers. The side-peaks surrounding the center-frequency originate from a non-zero parasitic magnetic field oriented in an arbitrary direction in the experimental environment, which is typically about 0.2-0.6 Gauss.[19,20] The peak at 40 MHz can often be observed, also in other samples[17] but the further investigation of the origin is not in the scope of this work. The relative intensity of ODMR was calculated as $(PL_{on} - PL_{off})/PL_{off}$ and it varies between 2%-4%, which is in good agreement with previous single spin studies.[2,15]

In conclusion, we have developed a scalable and efficient platform for collecting single silicon vacancy color center emission in silicon carbide through a free space objective or a lensed fiber interface, applicable for single photon generation and coherent control of spin-qubits. This platform presents an improvement in terms of scalability, device footprint and optical interface, while providing competitive collection efficiency. The approach can be extended to incorporate other color centers in bulk 3C-, 4H- and 6H-SiC samples.[1,3,13,21,22] In addition, silicon carbide is a biocompatible substrate[23] and can be directly interfaced with cells for potential magnetic[16,19,24] and temperature sensing applications,[4] whose proof-of-principle have recently been reported. Ideas for future development include increasing collection efficiency by fabricating taller nanopillars, isotopic purification of the sample for accessing qubits with longer spin coherence times, and integration with microwave control on chip. Another exciting direction is the investigation of recently suggested spin-photon interface protocols based on the $V_{Si}$ in SiC,[25] that requires the enhanced photon collection efficiency and the drastically improved ODMR signal of the $V_{Si}$ defects, which can be accessed at cryogenic temperatures.[26] There are several ways the irradiation could be modified to achieve higher yield of pillars with preferentially positioned single emitters. Control over the depth of the emitters would require irradiation with an ion beam with well-defined energy that preferentially generate color centers in the lower half of the pillars,[27] while lateral control can be achieved by the pre-masking of an array of apertures.[15] Another possibility to enhance the negatively charged[18] $V_{Si}$ yield at a desired position could lie in local doping control, as it was demonstrated e.g. with delta-doping for NV-centers in diamond[28]. Finally, focused ion beam irradiation can be applied to each fabricated nanopillar to generate color centers at a desired location, as has been recently demonstrated in diamond systems.[29]

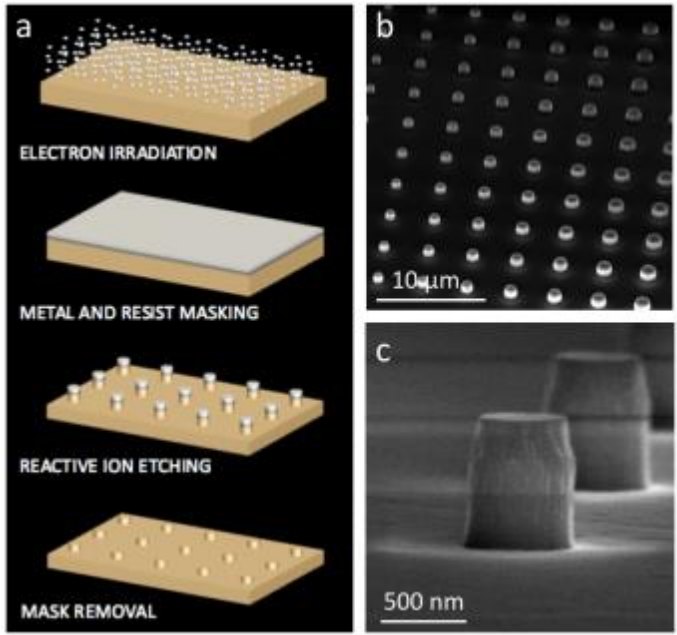

**Figure 1.** a) Lithographic process for fabrication of nanopillars in 4H-SiC. b) SEM images of fabricated array of nanopillars with variable diameters. c) SEM image of a 600 nm wide, 800 nm tall nanopillar.

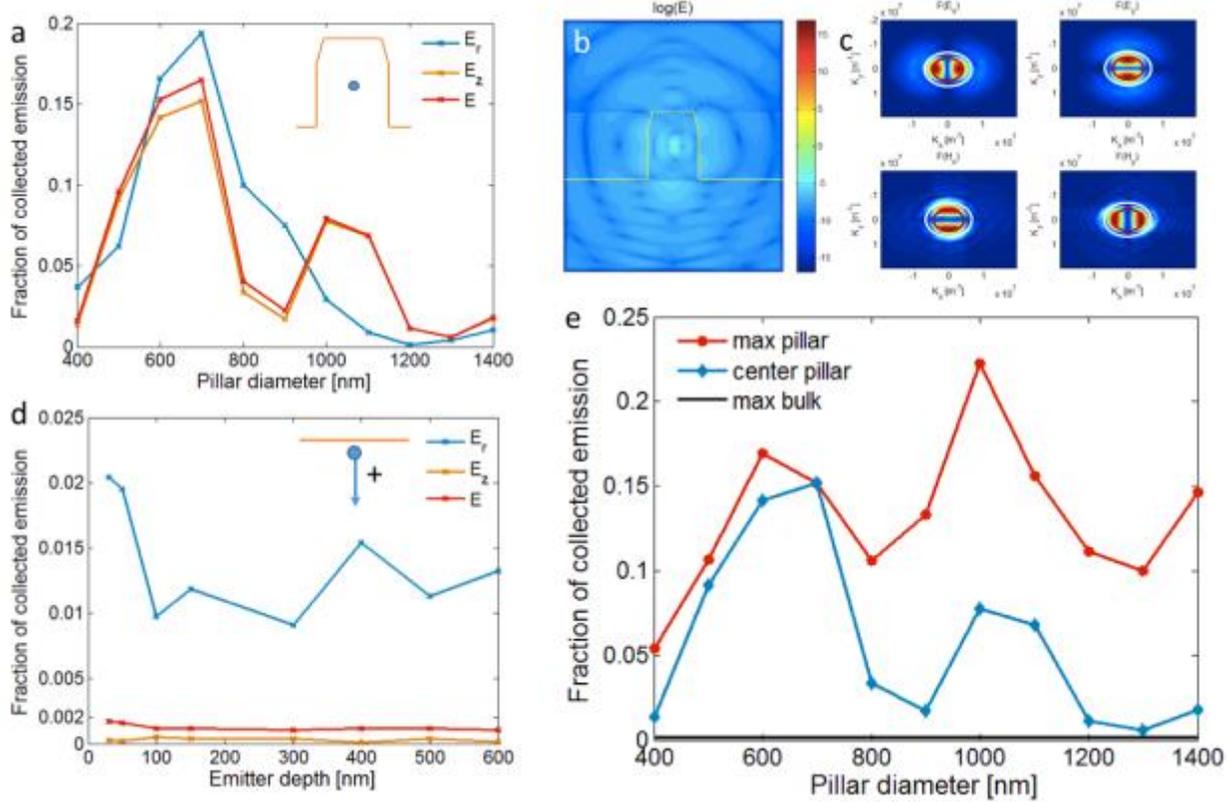

**Figure 2.** a) Collection efficiency from a dipole emitter positioned at the center of 800 nm tall SiC nanopillar (indicated by inset drawing), with 4° offset from z-axis (red), plotted with its vertical (yellow) and radial (blue) components. b) Finite Difference Time Domain model of electric field propagation from a dipole emitter placed in the center of a 600 nm wide nanopillar. c) Far field emission profile of b) with light line (white) and NA line (red). d) Color center depth dependence of collection efficiency from a dipole emitter in bulk SiC, with components marked as in a); inset drawing illustrates the direction of increasing depth. e) Diameter dependence of optimal (red) and centrally (blue) positioned color center in nanopillar, showing up to two orders of magnitude higher collection efficiency compared to bulk value of 0.0017.

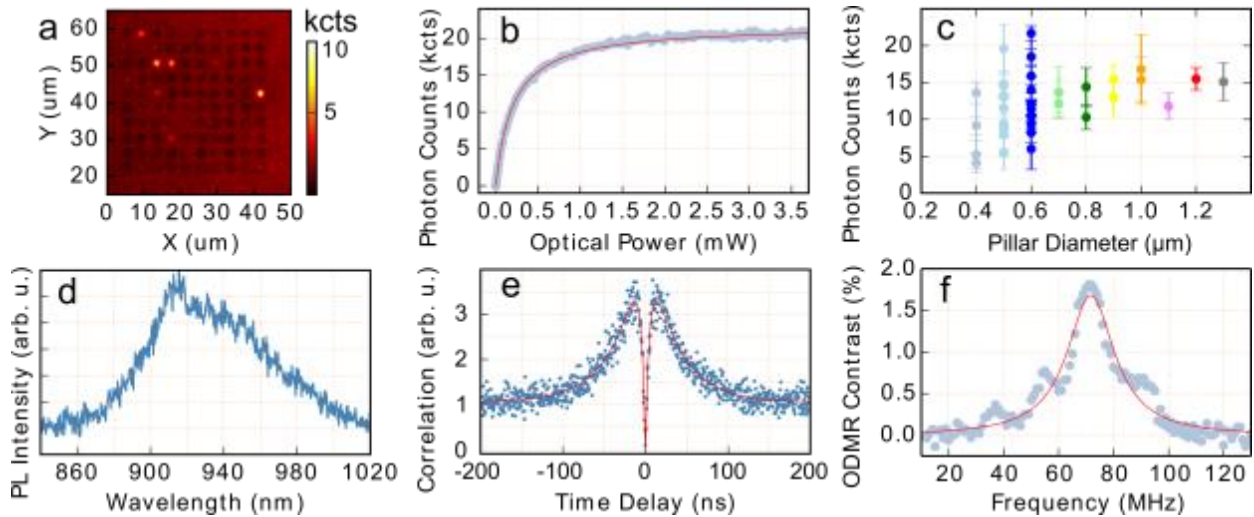

**Figure 3.** a) Photoluminescence intensity map of a block of pillars. b) PL intensity of a single $V_{Si}$ as a function of the optical power, saturating at 22 kcps. The red line shows a fit to the data. c) Diameter dependent collected count rate. d) Typical room temperature PL spectrum of a single $V_{Si}$ emitter in a nanopillar. e) Typical second order correlation function showing anti-bunching at zero time delay, a clear evidence for single photon emission. The red line shows a double exponential fit to the data (details in Supporting information). f) ODMR signal from a single $V_{Si}$ in a nanopillar. The red line shows a Lorentzian fit to the data.

AUTHOR INFORMATION

**Corresponding Author**

* marina.radulaski@stanford.edu

* m.widmann@physik.uni-stuttgart.de

**Author Contributions**

All authors have given approval to the final version of the manuscript. *These authors contributed equally.

**Funding Sources**

This work is supported by the National Science Foundation DMR Grant Number 1406028, by the Army Research Office under contract W911NF1310309, by the German Federal Ministry of Education and Research (BMBF) through the ERA.Net RUS Plus Project DIABASE, and by the JSPS KAKENHI (B) 26286047.

**Notes**

The authors declare no competing financial interest.

SUPPORTING INFORMATION

To analyze the influence of the position of V$_{Si}$ center in a nanopillar on collection efficiency we offset the excitation position vertically and horizontally away from the central position $r = 0$, $h = 400$ nm (Fig. S1). We analyze only *z*-polarized excitation, due to its dominant contribution to collection form the dipole aligned along c-axis forming angle $\theta = 4°$ with the vertical axis. Fig.

S1a shows that the collection is highest when the color center is placed in the bottom half of the pillar. This result can have practical implications to the depth of ion irradiation induced color centers in future approaches. Fig. S1b shows that the collection is highest close to the central axis of the pillar, which can be applied for masked irradiation which would target central positions along pillar diameter.

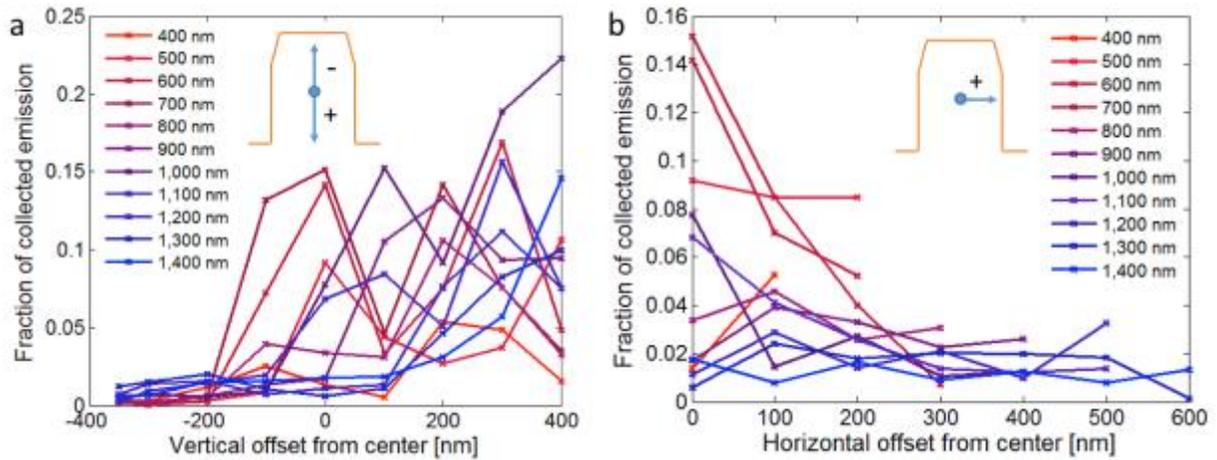

**Figure S1.** a) Collection efficiency dependence on vertical position along the central axis ($r = 0$), where negative values correspond to the top of the nanopillar, and positive ones to the bottom, plotted for various diameters. b) Collection efficiency dependence on horizontal position along the radius at the central heights ($h = 400$ nm), plotted for various diameters.

Using higher numerical aperture (NA) free space objective would additionally increase the collection efficiency from the nanopillars. Fig. S2 compares the collection into NA = 0.65 used in the experiment and NA = 0.95 that would achieve on average a two-fold increase in collection from the center of the nanopillar.

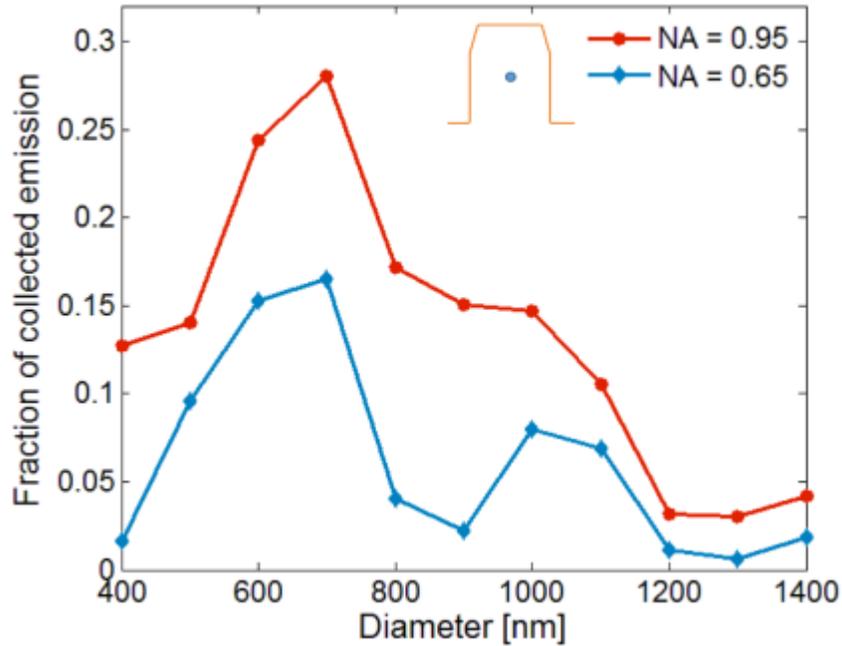

**Figure S2.** Collection efficiency from $V_{Si}$ at the center of the nanopillar ($r = 0$, $h = 400$ nm) into air objective with NA = 0.65 and NA = 0.95.

To characterize the potential nanofabrication influence on color center optical properties we fabricated a nanopillar system in a $V_{Si}$ rich 4H-SiC sample, where an ensemble of color center is hosted by each pillar. The pattern consisted of nanopillars and 40 μm × 40 μm square fields which provide intact regions, unaffected by etching chemistry. Fig. S3 shows the cryogenic temperature photoluminescence measurements of pillars and unetched region with same linewidth $\Delta\lambda = 0.05$ nm of the zero-phonon line V2 and increased collection efficiency.

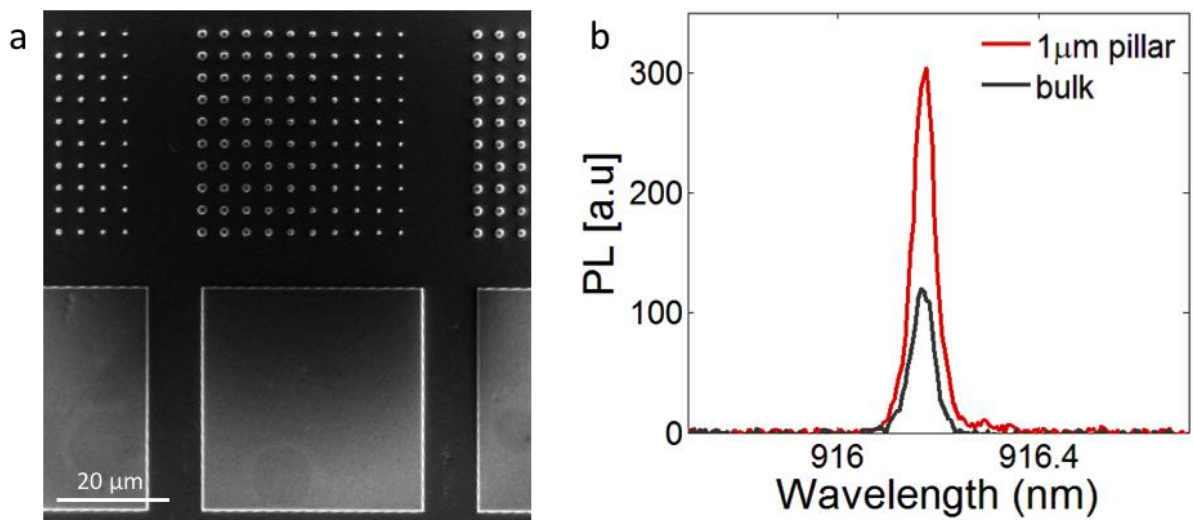

**Figure S3.** a) SEM image of a pattern with nanopillars and unetched bulk regions. b) Photoluminescence measurements at $T = 7$ K, showing same linewidth for nanopillars as in the bulk unetched region, as well as the increased collection efficiency.

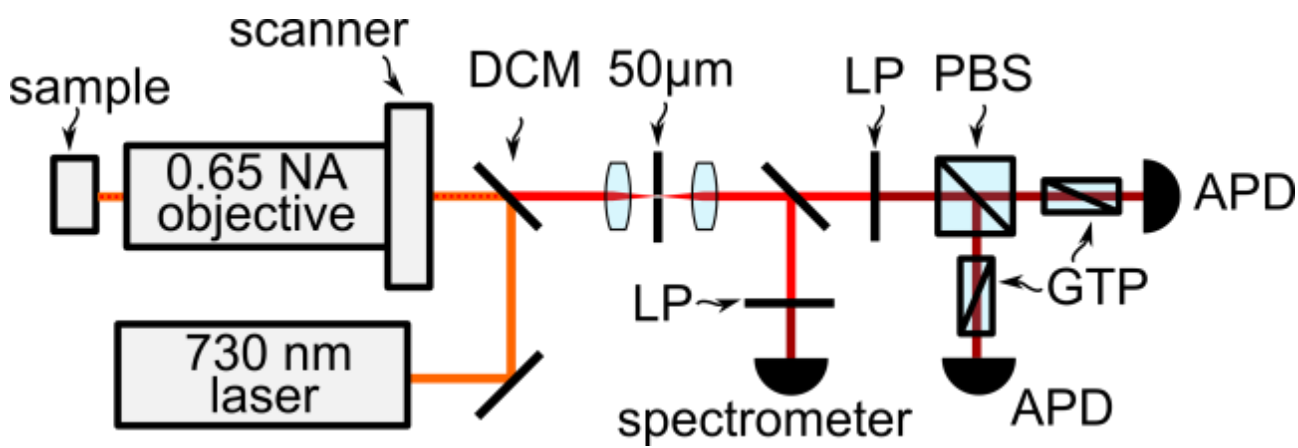

**Figure S4.** Experimental Setup. Orange line represents the 730 nm excitation beam, the red line shows the fluorescence, dark red is the filtered fluorescence. DCM indicates a dichroic mirror, long pass (LP) and polarizing beam-splitter (PBS), Glen-Taylor Polarizer (GTP) and single photon counting module as avalanche photo diode (APD).

Optical characterization of the sample is performed in the home-built confocal microscope shown in Fig. S4. A single mode laser (Openext) operating at 730 nm is sent over a dichroic mirror towards the objective (Zeiss, Plano-Achromat 40x/0.65NA), where it is focused onto the sample. The objective is mounted on a piezo-electric stage with a travel range in x-y-z of $(100 \times 100 \times 10)$ μm, respectively. Fluorescence from the emitters is collected by the same objective and passes through the dichroic mirror. Subsequently, the fluorescence is spatially filtered by focusing the light through a pinhole (50 μm), and spectrally filtered by a 905 nm long pass. The polarizing beam splitter (PBS), together with the avalanche photo diodes, form a Hanbury-Brown and Twiss (HBT) setup.[1] The additional Glen-Taylor calcite polarizers (GTP, Thorlabs GT10-B) are placed to suppress flash-back light from the APDs (Perkin Elmer SPCM-AQRH-15), which is caused by the avalanche of charge carriers accompanied by light emission.[2] Alternatively the light can be redirected before the 905 nm long pass filter to a spectrometer (Czerny-Turner type, Acton SpectraPro300i, grating: 300g mm$^{-1}$) equipped with a back-illuminated CCD camera (Princeton Instr., LN/CCD 1340/400 EHRB/l), filtered by a 800 nm long pass filter (Thorlabs, FELH0800).

Second order correlation measurements were obtained by analyzing photon-correlation using the HBT setup. The coincidences $c(\tau)$ obtained during the measurement time $T$ with a time resolution $\Delta t = 0.5$ ns is normalized to:

$$C_N(\tau) = \frac{c(\tau)}{N_1 N_2 \Delta t T}$$

Where $C_N(\tau)$ is the auto-correlation function, $N_1$ and $N_2$ are the photon count-rates on the APDs. To correct the signal $S$ for background light $B$, the auto-correlation function can be written as

$$g_c^2(\tau) = \frac{C_N(\tau) - (1-\rho^2)}{\rho^2},$$

where $\rho = S/(S + B)$. The measurements were performed under CW laser excitation operating at 730 nm with a power < 0.1 mW. The signal $S$ was found to have 1.5E3 photon counts on each APD and the background was 0.8E3 photon counts on each APD. Timing jitter of the APDs was found to a value of $\sigma = 0.5$ ns on the used setup,[3] hence a deconvolution of the photon-correlation measurement $c(\tau)$ with the instrument response function was not performed. The background-corrected second order correlation measurements $g_c^{(2)}(\tau)$ were fitted using a double exponential function:

$$g_c^{(2)}(\tau) = a_1 e^{-|\tau-\tau_0|/t_1} + a_2 e^{-|\tau-\tau_0|/t_2} + c,$$

with amplitudes $a_1, a_2$ and an offset value c. The value $\tau_0$ indicates a deviation from zero time-delay, and $t_1, t_2$ are time constants.

To drive the electron spins in the optically detected magnetic resonance (ODMR) measurement, radio-frequency electromagnetic field was generated by a signal generator (Rohde&Schwarz, SMIQ03B) which was subsequently amplified by a broadband RF amplifier (Minicircuits ZHL-42W) and applied to the defect by a 20 µm in diameter sized copper wire placed close to the defects.